\documentclass[12pt]{iopart}
\usepackage{iopams}
\usepackage[dvips]{graphicx}
\begin{document}
%%%%%%%%%%%%%%%%%%%%%%%%%%%%%%%%%%%%%%%%%%%%%%%%%%%%%%%%%
\title{Bulk viscosity and r-modes of neutron stars}
\author{Debarati Chatterjee $^{\rm (a,b)}$ and Debades 
Bandyopadhyay$^{\rm (a,b)}$}
\address{$^{\rm (a)}$Theory Division, Saha Institute of Nuclear Physics, 1/AF 
Bidhannagar, Calcutta 700 064, India}
\address{$^{\rm (b)}$Centre for Astroparticle Physics, Saha Institute of 
Nuclear Physics, 1/AF Bidhannagar, Calcutta 700 064, India}

\begin{abstract}
The bulk viscosity due 
to the non-leptonic process involving hyperons in $K^-$ condensed matter is 
discussed here. We find  that the bulk viscosity is modified in a 
superconducting phase. Further, we demonstrate how the exotic bulk viscosity 
coefficient influences $r$-modes of neutron stars which might be sources of 
detectable gravitational waves.
\end{abstract}

\section{Introduction}
Neutron stars have different classes of quasi normal modes (QNMs) \cite{Kok} 
depending on restoring forces acting on a perturbed fluid element.
Here we are interested in Coriolis restored fluid modes known as the inertial
$r$-modes. The $r$-modes become unstable due to gravitational radiation. It 
might be responsible for regulating the spin of newly born neutron stars as 
well as old, accreting neutron stars. It was shown that the leptonic and 
non-leptonic bulk viscosities might effectively damp the instability in 
different temperature regimes. The bulk viscosity coefficient due to
non-leptonic weak processes in neutron star matter involving 
hyperons \cite{Jon1,Jon2,Lin02,Han,Dal,Nar,DR1,DR2} , antikaon condensate 
\cite{DR3,DR4} and  quark matter \cite{Mad92,Mad00,Alf} was extensively
calculated by many authors. 

In this article, we discuss the hyperon bulk viscosity due to the 
non-leptonic process $n + p \rightleftharpoons p + \Lambda~$ 
in $K^-$ condensed matter and its
role on damping the $r$-mode instability. In section 2, we describe the 
composition and equation of state (EoS) of neutron star matter. Results of
hyperon bulk viscosity coefficients are explained in section 3. Section 4 gives
the summary.
 
\section{Composition and Equation of State}
Here we describe the composition and EoS of neutron star matter under going
a first order phase transition from hadronic to $K^-$ condensed matter. The 
constituents of charge neutral and $\beta$-equilibrated matter in both phases 
are neutrons ($n$), protons ($p$), $\Lambda$ hyperons, electrons, muons and 
also $K^-$ mesons in the condensed phase. The baryon-baryon interaction is 
mediated by the exchange
of $\sigma$, $\omega$ and $\rho$, $f_0$(975) (denoted hereafter as $\sigma^*$) 
and  $\phi$(1020) mesons and described by the  following Lagrangian density 
in Ref. \cite{Sch,Bog}. 

Similarly we adopt the Lagrangian density for (anti)kaons in the 
minimal coupling scheme as given by \cite{Gle98,Gle99,Pal,Bani1,Bani2}. 
The mixed phase is determined by the Gibbs phase rules and global charge 
neutrality and baryon number conservation \cite{Gle92}.

\begin{figure}[t]
\begin{center}
\includegraphics[height=8cm]{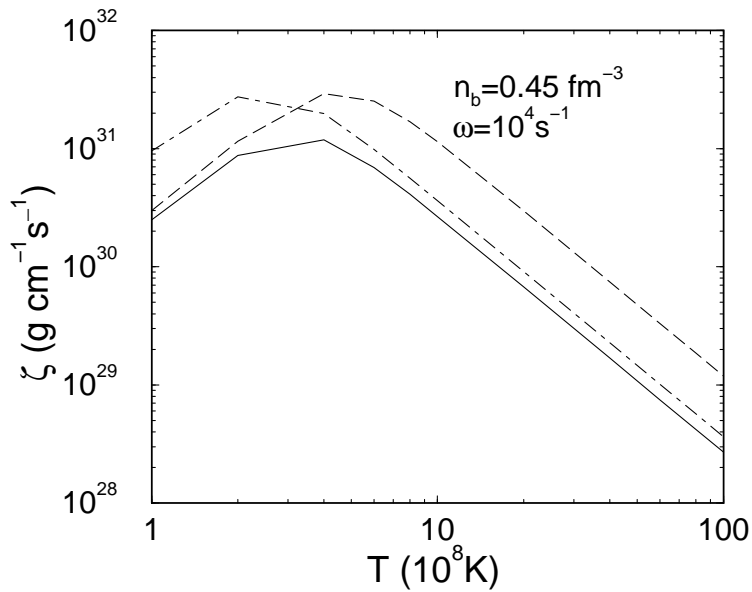} 
\caption{Bulk viscosity coefficient as a function of temperature with and 
without antikaon condensate for fixed values of baryon density ($n_b$) and 
angular velocity ($\omega$).}
\end{center}
\end{figure}

Nucleon-meson coupling constants which 
are determined by reproducing nuclear matter saturation
properties such as binding energy $E/B=-16.3$ MeV, baryon density $n_0=0.153$ 
fm$^{-3}$, asymmetry energy coefficient $a_{\rm asy}=32.5$ MeV, 
incompressibility $K=300$ MeV and effective nucleon mass $m^*_N/m_N = 0.70$, 
are taken from Ref.\cite{Gle91}. Further kaon-meson 
and  hyperon-meson coupling constants are obtained from Ref. 
\cite{DR4,Sch}.
In this calculation, we adopt the value of antikaon optical potential depth 
at normal nuclear matter density as $U_{\bar K}(n_0) = -160$ MeV. 

When there is no first order phase
transition from hadronic to $K^-$ condensed matter, $\Lambda$ hyperons 
appear at density 2.25$n_0$. On the other hand, the early onset of 
$K^-$ condensate in a first order antikaon condensation, delays the appearance
of $\Lambda$ hyperons. In our calculation, $K^-$ condensation begins at 
2.23$n_0$ and the mixed phase ends at 4.1$n_0$ whereas $\Lambda$ hyperons 
appear at 2.51$n_0$. Further we calculate energy density and pressure 
corresponding to above mentioned compositions. 

\section{Hyperon bulk viscosity coefficient and suppression of $r$-modes}

The real part of bulk viscosity coefficient due to the
non-leptonic process $n + p \rightleftharpoons p + \Lambda$ is given 
by \cite{Lin02,Lan}, 
\begin{equation}
\zeta = \frac {P(\gamma_{\infty} - \gamma_0)\tau}{1 + {(\omega\tau)}^2}~.
\end{equation}
The difference of infinite and zero frequency adiabatic indices is,
\begin{equation}
\gamma_{\infty} - \gamma_0 = - \frac {n_b^2}{P} \frac {\partial P} 
{\partial n_n} \frac {d{\bar x}_n} {dn_b}~,
\end{equation}
where $\bar x_n = \frac {n_n}{n_b}$ is the neutron fraction in the equilibrium 
state.
The relaxation time ($\tau$) for the non-leptonic process in the j-th 
(= hadronic / $K^-$ condensed) phase 
is given by \cite{Lin02,Nar,DR4}
\begin{equation}
\frac {1}{\tau} = \frac {{(kT)}^2}{192{\pi}^3} {p_{\Lambda}}
{<{{|M_{\Lambda}|}^2}>} \frac {\delta \mu}{\delta{n_n^j}}~,
\end{equation}
where $p_{\Lambda}$ is the Fermi momentum for $\Lambda$ hyperons and 
$<{|M_{\Lambda}|}^2>$ is the angle averaged matrix element squared in the
corresponding phase. 

Temperature dependence of hyperon bulk viscosities due to the non-leptonic 
process $n + p \rightleftharpoons p + \Lambda$ with and without $K^-$ 
condensate are shown at $n_b = 0.45 fm^{-3}$ and $\omega = 10^4 s^{-1}$ in
Figure 1. The solid line represents the hyperon bulk viscosity in the absence of
antikaon condensate. The dashed and dashed-dotted lines correspond to the
hyperon bulk viscosities in the hadronic and antikaon condensed phases of the
mixed phase in a first order hadronic to $K^-$ condensed phase transition 
\cite{DR4}. The hyperon bulk viscosity in the absence of $K^-$ condensate is
suppressed compared with hyperon bulk viscosities in the presence of antikaon 
condensate. Further we note that the hyperon bulk viscosity coefficient in the 
condensed phase is smaller than that of the hadronic phase above 
3 $\times 10^{8}$K. This is also the temperature where the inversion of the
temperature dependence of hyperon bulk viscosity in $K^-$ condensate happens.
This feature is also found in the case without the condensate. 

\begin{figure}[t]
\begin{center}
\includegraphics[height=8cm]{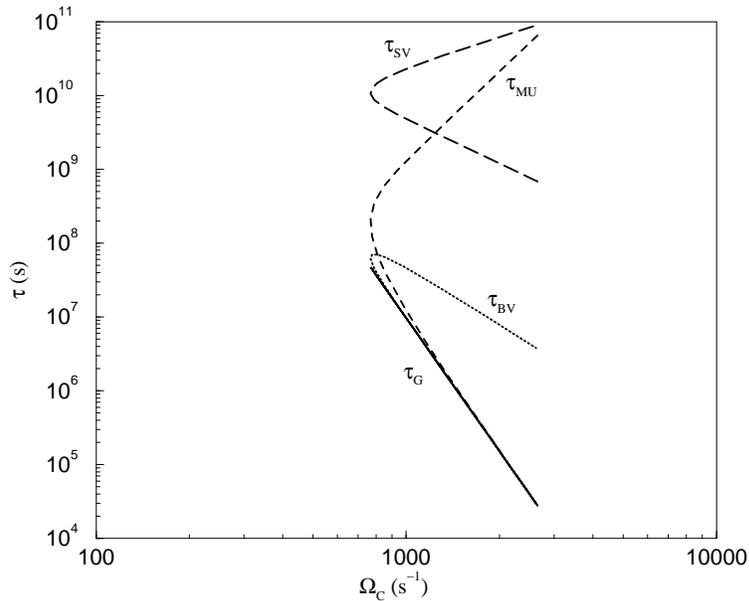} 
\caption{Different damping time scales and the growth time scale due to   
gravitational radiation are shown as a function critical angular velocity  
($\Omega_C$).}
\end{center}
\end{figure}

We investigate the r-modes of a neutron star having gravitational mass 
1.6 M$_{\odot}$ corresponding to central baryon density 3.5$n_0$ and rotating
at an angular velocity 2652 $s^{-1}$. We calculate damping time scales 
corresponding to hyperon bulk viscosity ($\tau_{BV}$), modified Urca bulk
viscosity involving nucleons ($\tau_{MU}$), shear viscosity ($\tau_{SV}$) and
the growth time scale of gravitational radiation ($\tau_{G}$) using the energy
density and bulk viscosity profiles \cite{DR4}.
We compute the critical angular velocities ($\Omega_C$) as a function 
temperature solving
$ - {\frac {1}{\tau_{GR}}} + {\frac {1}{\tau_B}} + 
{\frac {1}{\tau_U}} + {\frac {1}{\tau_{SV}}} = 0~$.
Various damping time scales are exhibited as a function of critical angular 
velocity in Figure 2.
It is noted that the $r$-mode instability of the neutron star is damped when 
the growth time scale due to
gravitational radiation is comparable with damping time scales $\tau_{BV}$ 
and $\tau_{MU}$. We find that the hyperon bulk viscosity damps the instability
at $T = 4 \times 10^{9}$ and below whereas it is the modified Urca bulk 
viscosity which effectively suppresses the $r$-mode instability above this 
temperature.      

\section{Summary}

We have studied hyperon bulk viscosity in the presence of $K^-$ condensate. 
We find the inversion of the temperature dependence of hyperon bulk viscosity 
in our calculation. Though the hyperon bulk viscosity is suppressed in $K^-$
condensed matter than that in the hadronic phase, it still effectively damps 
the $r$-mode instability.


\section*{References}

\begin{thebibliography}{99}
\bibitem {Kok} Andersson N and Kokkotas K D 2001 {\it Int. J. Mod. Phys.} D 
{\bf 10} 381
\bibitem {Jon1} Jones P B 2001 {\it Phys. Rev. Lett.} {\bf 86} 1384 
\bibitem {Jon2} Jones P B 2001 {\it Phys. Rev.} D {\bf 64} 084003 
\bibitem {Lin02} Lindblom L and Owen B J 2002 {\it Phys. Rev.} D {\bf 65} 
063006 
\bibitem {Han} Haensel P, Levenfish K P and
Yakovlev D G 2002 {\it Astron. Astrophys.} {\bf 381} 1080
\bibitem {Dal} van Dalen E N E and Dieperink A E L 2002 {\it Phys. Rev.} D 
{\bf 65}, 063006
\bibitem {Nar} Nayyar M and Owen B J 2006 {\it Phys. Rev.} D {\bf 73} 084001 
\bibitem {DR1} Chatterjee D and Bandyopadhyay D 2006 {\it Phys. Rev.} D 
{\bf 74} 023003
\bibitem {DR2} Chatterjee D and Bandyopadhyay D 2007 {\it Astrophys. Space Sc.}
{\bf 308} 451
\bibitem {DR3} Chatterjee D and Bandyopadhyay D 2007 {\it Phys. Rev.} D 
{\bf 75} 123006
\bibitem {DR4} Chatterjee D and Bandyopadhyay D 2008 {\it ApJ} {\bf 680} 686 
\bibitem {Mad92} Madsen J 1992 {\it Phys. Rev.} D {\bf 46} 3290
\bibitem {Mad00} Madsen J 2000 {\it Phys. Rev. Lett.} {\bf 85} 10
\bibitem {Alf} Alford M G, Rajagopal K, Schaefer T and Schmitt A 2008
{\it Preprint} arXiv:0709.4635
\bibitem{Sch} Schaffner J and Mishustin I N 1996 
{\it Phys. Rev.} C {\bf 53} 1416 
\bibitem{Bog} Boguta J and Bodmer A R 1977 {\it Nucl. Phys.} A {\bf 292} 413 
\bibitem{Gle98} Glendenning N K and Schaffner-Bielich J 1998 
{\it Phys. Rev. Lett.} {\bf 81} 4564
\bibitem{Gle99} Glendenning N K and Schaffner-Bielich J 1999 {\it Phys. Rev.} C
{\bf 60} 025803
\bibitem {Pal} Pal S, Bandyopadhyay D and Greiner W 2000 {\it Nucl. Phys.} A 
{\bf 674} 553
\bibitem{Bani1} Banik S and Bandyopadhyay D 2001 {\it Phys. Rev.} C {\bf 63}
035802
\bibitem{Bani2} Banik S and Bandyopadhyay D 2001 {\it Phys. Rev.} C {\bf 64} 
055805
\bibitem{Gle92} Glendenning N K 1992 {\it Phys. Rev.} D {\bf 46} 1274
\bibitem{Gle91} Glendenning N K and Moszkowski S A 1991 {\it Phys. Rev. Lett.}
{\bf 67} 2414
\bibitem{Lan} Landau L D and Lifshitz E M
1999 Fluid Mechanics (Oxford:Butterworth-Heinemann) 
\end{thebibliography}
\end{document}